\documentclass[aps,prd,superscriptaddress,showpacs,twocolumn,eqsecnum,notitlepage,nofootinbib,10pt]{revtex4-1}

\usepackage{color}
\usepackage[usenames,dvipsnames,svgnames,table]{xcolor}
\usepackage{amsmath}
\usepackage{amsfonts}
\usepackage{amssymb}
\usepackage{amsthm}
\usepackage{mathrsfs}
\usepackage{eucal}
\usepackage{MnSymbol}
\usepackage{mathtools}
\usepackage{tensor}
\usepackage{subfigure}
\usepackage{url}

\usepackage[citecolor=Blue,colorlinks=true,linkcolor=Blue,urlcolor=Blue]{hyperref}

\newcommand{\mbs}{\boldsymbol}
\newcommand{\mca}{\CMcal}

\newcommand{\msf}{\mathsf}

\newcommand{\mbb}{\mathbb}

\newcommand{\tit}{\textit}

\newcommand{\txt}{\text}
\newcommand{\pde}{\partial}

\newcommand{\nal}{\natural}

\newcommand{\dms}{\diamondsuit}

\newcommand{\ima}{\msf{i}}
\newcommand{\equi}{\equiv}
\newcommand{\eqre}{\triplesim}
\newcommand{\eref}{\eqref}
\newcommand{\lbel}{\label}

\newcommand{\sig}{\txt{sig}}

\newcommand{\exd}{\txt{d}}

\begin{document}

\title{Singularity resolution from polymer quantum matter}

\author{Andreas Kreienbuehl}
\email{a.kreienbuehl@hef.ru.nl}
\affiliation{Theoretical High Energy Physics, Radboud University, Mailbox 79, P.O. Box 9010, 6500 GL Nijmegen, The Netherlands.}

\author{Tomasz Paw{\l}owski}
\email{tpawlow@fuw.edu.pl}
\affiliation{Departamento de Ciencias F\'isicas, Facultad de Ciencias Exactas, Universidad Andres Bello, Av.~Rep\'ublica 220,  Santiago de Chile.}
\affiliation{Katedra Metod Matematycznych Fizyki, Universytet Warszawski, ul. Ho\.za 74, 00-681 Warszawa, Poland.}


\begin{abstract}
We study the polymeric nature of quantum matter fields using the example of a Friedmann-Lema\^itre-Robertson-Walker universe sourced by a minimally coupled massless scalar field. The model is treated in the symmetry reduced regime via deparametrization techniques, with the scale factor playing the role of time. 
Subsequently, the remaining dynamic degrees of freedom corresponding to the matter are polymer quantized. The analysis of the resulting genuine quantum dynamic shows that the big bang singularity is resolved, although with   
the form of the resolution differing significantly from that in the models with matter clocks: dynamically, the singularity is made passable rather than avoided. Furthermore, this analysis exposes crucial limitations to the so-called effective dynamic in loop quantum cosmology when applied outside of the most basic isotropic settings.
\end{abstract}

\maketitle

\section{Introduction}

Einstein's theory of general relativity (GR) successfully describes gravitational phenomena, predicting with high precision all large scale observations made to date. It is however expected to fail in the ultraviolet regime due to the quantum nature of the reality at the Planck energy scale. To obtain accurate predictions for such situations one has to resort to quantum gravity (QG).

Despite many attempts \cite{loll:98,ashtekar:04,kiefer:07,rovelli:07,thiemann:08,bojowald:11}, no general, complete, and working (quantitatively) formulation of QG exists. In particular, in the context of the canonical quantization programs, so far it was possible to complete the quantization program only in certain situations, where gravity is coupled to specific matter fields (for irrotational dust see \cite{husain:11a,hp-dust-det} and also the earlier, well defined implicit constructions in \cite{gt-dust,dgkl-gq}). In order to generalize these frameworks (or complete the alternative approaches) it is crucial to first study in detail the simplified mini- and midisuperspace settings.

In this paper we consider the minisuperspace model, which represents a Fried\-mann-Le\-ma\^it\-re-Rob\-ert\-son-Walk\-er (FLRW) universe -- an isotropic and flat spacetime admitting a massless scalar field as source. This model is widely used as a testing ground for QG methods and is at the same time of particular interest in cosmology.

In the context of QG the model has been studied in detail using tools of loop quantum cosmology (LQC -- see the references in the second paragraph and \cite{b-livrev,a-lqc-overview,*a-lqc-intro,ashtekar:11,mm-intro,bcmb-rev,bojowald:01,ashtekar:06a}). In LQC, an application of the polymer quantization \cite{n-outside,halvorson:03,ashtekar:03a,t-inside,al-rev} to the geometric degrees of freedom results in a dynamical singularity resolution 
\cite{ashtekar:06a}, whereby the big bang is replaced by a big bounce. This result was later confirmed (at the genuine quantum level) for different matter fields, in particular 
the Maxwell field \cite{ppwe-rad} and dust \cite{husain:11b}.

The above-mentioned big bounce result was however obtained via a somewhat ``hybrid'' approach: the geometry is quantized via loop techniques, while the matter (the scalar field) is treated by methods of standard quantum mechanics (Schr\"odinger representation).\footnote{Recently, the results of \cite{ashtekar:06a} were confirmed \cite{ddl-scalar} through the analysis of the same model, where both the geometric and the matter degrees of freedom are quantized via polymer techniques. This work used one of several possible in this context loop quantization schemes (see \cite{halvorson:03} and the discussion in Sec.\;\ref{s:11}).} Furthermore, the system was analyzed by methods dedicated to theories with a time reparametrization freedom. Namely, the evolution was implicitly defined by means of the formalism of partial observables \cite{obs-r,*obs-d}.
The above approach was also applied outside of the isotropic settings, both for homogeneous models (like various Bianchi models \cite{awe-b1,*awe-b2,*awe-b9}) and inhomogeneous spacetimes (in particular Gowdy models \cite{mbmbmm-gmat}) as well as in the context of perturbation theory about the cosmological sectors \cite{Agullo:2012sh,*Agullo:2012fc,*Agullo:2013ai,FernandezMendez:2012sr,*FernandezMendez:2012vi,Dapor:2013pka}.\footnote{The list of references given here contains only selected examples representing the current state of development for each model. For a more complete list we refer the reader to \cite{ashtekar:11,bcmb-rev}.} The results for the inhomogeneous settings are however based on heuristic methods and the dynamic is not systematically investigated.

An alternative approach is presented in \cite{kreienbuehl:09} (see also \cite{blyth:75}), where the investigated model is the one considered here but including a nonvanishing cosmological constant. The analysis is carried out in the context of quantum geometrodynamic, which is based on the standard Schr\"odinger quantization of the metric Hamiltonian formulation of Arnowitt, Deser, and Misner (ADM). More specifically, the system is treated via the so-called deparametrization technique (for an example starting from the full theory see \cite{gt-dust}): one of the dynamical variables -- in this case the scale factor -- is selected as clock at the classical level, after which the system can be quantized and regarded as freely evolving with respect to this clock. The obtained results are (by the majority) consistent with those of the studies of the same systems in the framework of geometrodynamic in \cite{ashtekar:06c,bp-negL,pa-posL,*kp-posL}, where the partial variable formalism was applied \cite{obs-r,*obs-d}.

A consistent treatment requires the quantization of the geometry and the matter in the same way. In the context of LQC the intermediate step towards this goal is the analysis of a loop quantized scalar field coupled to gravity quantized via standard techniques. This step is necessary to identify the physical effects arising specifically due to the polymer nature of matter.

In full loop quantum gravity (LQG) a consistent quantization of the scalar field was proposed in \cite{klb-scalar,*klo-scalar}, where one of two possible (and inequivalent) implementations of the polymer representation \cite{halvorson:03} was used. The same choice was later made in \cite{ddl-scalar} to derive the symmetry reduced description and to determine the LQC dynamic of (this form of) the polymer scalar field.

The alternative (in a certain sense dual to the above) consistent quantization prescription was applied in \cite{hossain:10a}, where again the FLRW isotropic universe is investigated by scalar quantum mechanics on a classical cosmological background. Elements of a semiclassical analysis led to the construction of an effective approximation of the dynamic, of which the study showed that the big bang singularity is replaced by a past-eternal de Sitter phase (``eternal inflation'') with graceful exit.

The mathematical formalism characteristic to this prescription was later successfully extended to the inhomogeneous setting in the context of quantum field theory on Minkowski space and on a cosmological background \cite{hossain:09,*hossain:10b,*hossain:10c,*husain:10}. Both of these extensions were built via Bojowald's lattice refinement techniques \cite{bck-latref}.

\tit{Since the nature of the studies mentioned just now is semi-heuristic, a comparison with the genuine quantum dynamic is indispensable. This is exactly the goal of the work presented here.} To provide the precise quantum theory we first perform a deparametrization analogous to the one in \cite{kreienbuehl:09}, choosing the scale factor cubed as clock. Then, we quantize the scalar field via loop techniques, applying the prescription originally provided in \cite{hossain:10a}.

In our work we focus on the precise construction of the quantum model, that is in particular the correct definition of the Hilbert space, and on the analysis of the physical consequences: the dynamic, the existence of a semiclassical sector, and the correct GR limit. Surprisingly, the requirement of the latter will have a critical impact on the form of the Hilbert space and, consequently, on the domain of applicability of the heuristic construction of the so-called effective dynamic from LQC.

The numerical analysis of the dynamic shows that there is no quantum big bang. However, instead of bouncing back, the quantum state (the wave packet) transits deterministically through the point marking the singularity in GR. The quantum evolution picture appearing here resembles thus the one advertised in the ``early LQC epoch'' \cite{b-livrev}.\footnote{We note that the early results were derived for a different system, where the geometry instead of the matter was polymeric.} Consequently, this work (see also the results in \cite{ddl-scalar}) suggests that the big \tit{bounce} is an effect arising solely due to the polymeric (discrete) quantum nature of the \emph{geometry}.

The paper is structured as follows: in Sec.\;\ref{s:10} we introduce the details of the classical FLRW model that we analyze. Then we proceed in Sec.\;\ref{s:11} with the construction of the precise quantum theory. Finally, in Sec.\;\ref{s:0} we analyze the physical results and conclude in Sec.\;\ref{s:12} with a general discussion.

In our studies we select the natural units $c\equi\hslash\equi1$ and introduce the abbreviation $L\equi(12\pi G_{\txt{N}})^{1/2}$ for a length scale. Later in the paper we further restrict our attention to the case $L=1$ corresponding to a form of Planck units.

\section{\lbel{s:10}Classical theory}

In this paper we focus on the case of an isotropic and flat FLRW universe with a minimally coupled massless scalar field $\phi\equi\phi(T)\in\mbb{R}$ as source. The metric tensor of such a universe can be expressed as
\begin{equation}\label{eq:g-dec}
	g\equi-(N\exd T)^2+a^2\delta_{ij}\exd X^i\exd X^j,
\end{equation}
where $N\equi N(T)\in\mbb{R}^+$ is the lapse function and $a\equi a(T)\in\mbb{R}$ is the scale factor (we are working in an ``extended'' minisuperspace \cite{carmeli:1970}). Since the ``symmetric criticality principle'' \cite{kiefer:07,palais:79} is valid in the present situation, the canonical action
\begin{equation}\lbel{e:1000}
	A=\int_{T_{\txt{I}}}^{T_{\txt{F}}}(P_a\dot{a}+P_{\phi}\dot{\phi}-H[N])\;\exd T
\end{equation}
can be directly and conveniently derived from the reduced Einstein-Hilbert action. The canonical momenta appearing in \eref{e:1000} are
\begin{subequations}
\begin{gather}
	P_a=\frac{9V_{\txt{C}}}{L^2}\frac{|a|\dot{a}}{N},\qquad P_{\phi}=V_{\txt{C}}\frac{|a|^3\dot{\phi}}{N},\\
	\{a,P_a\}=1,\qquad\{\phi,P_{\phi}\}=1,
\end{gather}
\end{subequations}
whereas the scalar Hamiltonian constraint takes the form
\begin{equation}\lbel{e:1004}
	H[N]=N\frac{V_{\txt{C}}|a|^3}{2L ^2}\left[-\left(\frac{L ^2P_a}{3V_{\txt{C}}a^2}\right)^2+\left(\frac{L P_{\phi}}{V_{\txt{C}}a^3}\right)^2\right].
\end{equation}
Note that to arrive at the form \eref{e:1000} of the action we had to first introduce the 1+3 splitting $\mca{M}\simeq\mbb{R}\times\mca{N}$, where for the flat universe (considered here) $\mca{N}=\mbb{R}^3$. Due to the noncompactness of the spatial slices we had to then introduce in the process of deriving \eref{e:1000} an infrared regulator -- a cube or ``cell'' $\mca{V}\subset\mbb{R}^3$ of finite size (see \cite{ashtekar:06b,ashtekar:06c} and the discussion in \cite{cm-cell}). Its physical volume is $V=V_{\txt{C}}|a|^3$, where $V_{\txt{C}}\equi\int_{\mca{V}}\exd^3X$ is the comoving coordinate volume of $\mca{V}$. 

This infrared regulating step introduces an additional complication into the treatment, as one has to make sure that the resulting model has a well defined (unambiguous) regulator removal limit. 
The classical FLRW theory is invariant under the rescaling $\vec{X}\mapsto\zeta\vec{X}$, $\zeta\in\mbb{R}^+$, which increases $V$ by a factor of $\zeta^3$ (an \tit{active} diffeomorphism). This invariance is a natural requirement for the description of the model to remain well defined when the regulator is removed \cite{acs-rob}. We stress that this requirement is however by no means sufficient in the quantum theory (see in particular \cite{cm-cell}). Furthermore, in the quantum theory the $\zeta$-invariance is not given trivially \cite{ashtekar:06b,ashtekar:06c,ashtekar:11} and hence imposing it as a condition for consistency affects the choice of the canonical variables for the quantization [see \eref{e:101}, \eref{e:1001}, and the paragraph prior to them].

The \tit{first}-class \cite{dirac:64,hanson:76,henneaux:92} Hamiltonian constraint $H[N]\approx0$ generates \tit{infinitesimal} transformations of $T$. As can be seen from \eref{eq:g-dec}, there also is the possibility of a reparametrization of the scale factor $a$ and the Euclidean metric $\delta$ by an $\eta\in\mbb{R}^{\pm}$ such that  $a\mapsto a/\eta$ and $\delta\mapsto\eta^2\delta$, respectively. This residual $\eta$-symmetry corresponds to the freedom of fixing the coordinate scale (a \tit{passive} diffeomorphism by $|\eta|$) and orientation $\mbs{(}$a ``large gauge transformation'' \cite{ashtekar:06b,ashtekar:06a,ashtekar:06c,ashtekar:07,ashtekar:11} by $\sig(\eta)$$\mbs{)}$. Just like it is required for the $\zeta$-transformation, the physics has to be invariant under an $\eta$-transformation. Among the $\eta$-invariant quantities are the action $A$, the volume $V$, and ratios of the scale factor such as the Hubble parameter $h\equi\dot{a}/(Na)$.

In the next step we fix the time reparametrization freedom and time orientation by implementing the \tit{second}-class \cite{dirac:64,hanson:76,henneaux:92} gauge
\begin{equation}\lbel{e:73}
	G\equi T-\frac{V_{\txt{C}}a^3}{L ^2}.
\end{equation}
Given the equation of motion
\begin{equation}\lbel{e:31}
	\dot{a}=\{a,H[N]\}=-N\frac{L ^2P_a}{9V_{\txt{C}}|a|},
\end{equation}
the form of $G$ implies in particular that $P_a$ is negative. This and the form of the constraint $H[N]$ then lead to the reduced canonical action
\begin{equation}\lbel{e:72}
	A=\int_{T_{\txt{I}}}^{T_{\txt{F}}}(P_{\phi}\dot{\phi}-H_{\txt{R}})\;\exd T,
\end{equation}
where again $\{\phi,P_{\phi}\}=1$. The reduced Hamiltonian takes the form
\begin{equation}\lbel{e:70}
	H_{\txt{R}}\equi-P_T=-\frac{L ^2P_a}{3V_{\txt{C}}a^2}=\left|\frac{L ^2P_a}{3V_{\txt{C}}a^2}\right|=\left|\frac{P_{\phi}}{L T}\right|.
\end{equation}
Finally, the consistency condition
\begin{equation}
	\pde_TG+\{G,H[N]\}=0
\end{equation}
uniquely determines the lapse function
\begin{equation}
	N=\frac{L }{|P_{\phi}|}.
\end{equation}

We emphasize that the time-gauge $G$ becomes $T-\sig(\eta)V_{\txt{C}}a^3/L ^2$ under an $\eta$-transformation. This means that the orientation of $a$ relative to $T$ changes if $\eta\in\mbb{R}^-$. Replacing $T$ by $-T$ has no impact on the space of solutions to the Wheeler-DeWitt equation resulting from \eref{e:1004} and amounts to a ``time-reversal'' operation \cite{catren:06}. However, the \tit{reduced} classical formalism derived here is the result of singling out one of $\pm T$ [see \eref{e:73}] so that it is not time-reversal invariant. If \eref{e:73} defines a future-directed clock, the past orientation would be given by the gauge constraint $G=T+V_{\txt{C}}a^3/L ^2$. These considerations are relevant in the construction of the initial state for the quantum evolution [see \eref{e:4141} and the paragraph containing this equation].

The reduced canonical formalism we constructed just now has the deficiency of explicitly depending on the infrared regulator $\mca{V}$ since, according to \eref{e:73}, the clock variable $T$ scales like $\zeta^3$. This would make the removal of the infrared regulator $\mca{V}$ from the resulting quantum theory a rather tedious task. As the initial step in addressing this problem we replace $T$ by the dimensionless variable $t\equi T/|T^{\dms}|$, where $T^{\dms}\in\mbb{R}^{\pm}$ is some fixed but otherwise arbitrary reference value. Furthermore, since $P_{\phi}$ also scales like $\zeta^3$, we analogously define $p_{\phi}\equi P_{\phi}/|P_{\phi}^{\dms}|$ with $P_{\phi}^{\dms}\in\mbb{R}^{\pm}$ being a fixed reference value for $P_{\phi}$. Technically, the replacement of $T$ by $t$ can be brought about by a change of the integration variable in \eref{e:72}, whereas the replacement of $P_{\phi}$ by $p_{\phi}$ is realized by an ``extended canonical'' or ``scale transformation'' \cite{goldstein:01}. Altogether, this procedure yields the canonical action
\begin{equation}\lbel{e:101}
	A=\frac{|P_{\phi}^{\dms}|}{L }\int_{|T^{\dms}|t_{\txt{I}}}^{|T^{\dms}|t_{\txt{F}}}(L p_{\phi}\phi'-H_{\txt{S}})\;\exd t,
\end{equation}
where $\phi'\equi d\phi/dt$, $\{\phi,p_{\phi}\}=1/L $, and
\begin{equation}\lbel{e:1001}
	H_{\txt{S}}\equi\left|\frac{p_{\phi}}{t}\right|,\quad\left|\frac{p_{\phi}}{t}\right|\exd t=\frac{L }{|P_{\phi}^{\dms}|}H_{\txt{R}}\;\exd T,
\end{equation}
is the Hamiltonian related to $H_{\txt{R}}$ by a \tit{scaling}. We stress that $t$ and $p_{\phi}$ are dimensionless variables.

At this point it is necessary to mention that the above modification does not yet completely remove the dependence of the theory on the infrared regulator. Indeed, while the constants $T^{\dms}$ and $P^{\dms}_{\phi}$ are fixed, no particular value of $\mca{V}$ can be distinguished on physical grounds. In consequence, the particular ``physical'' universe is represented by classes of solutions rather than by single ones. This nonuniqueness can be easily shown at the level of specifying the initial data. There, the single universe regulated by different cells $\mca{V}$ will correspond to the entire set (equivalence class) of the initial data at a chosen initial time $t^{\star}$. This dependence will propagate through to the quantum theory.

For initial $t^{\star},p_{\phi}^{\star}\in\mbb{R}^{\pm}$ the solutions to Hamilton's equations of motion derived from \eref{e:101} and \eref{e:1001} are
\begin{equation}\lbel{e:14}
	p_{\phi}(t)=p_{\phi}^{\star},\quad\phi(t)=\phi^{\star}+\frac{\sig(tp_{\phi}^{\star})}{L }\ln\left(\left|\frac{t}{t^{\star}}\right|\right).
\end{equation}
Therefore, the lapse function $N=|T^{\dms}|/|P_{\phi}^{\dms}p_{\phi}|$ is a constant [positive because of the chosen gauge constraint \eref{e:73}]. The canonical representation of the spacetime Ricci scalar takes then the form $R=-6h^2$, which in turn implies
\begin{equation}\lbel{e:1003}
	-\frac{3R(t)}{2R^{\dms}}=\left[\frac{p_{\phi}(t)}{t}\right]^2=\left(\frac{p_{\phi}^{\star}}{t}\right)^2,\quad R^{\dms}\equi \left(\frac{P_{\phi}^{\dms}}{L T^{\dms}}\right)^2.
\end{equation}
The form of the reference value $R^{\dms}$ suggests the natural and simplifying choice $|P_{\phi}^{\dms}|=|T^{\dms}|$, which corresponds to fixing the initial curvature value to be $R(t^\star)=-(2/3)[p_{\phi}^{\star}/(L t^{\star})]^2$. The value of $|P_{\phi}^{\dms}|$ relative to $|T^{\dms}|$ is now fixed but the implicit $\xi$-dependence discussed earlier is still present.

In order to simplify the expressions, from now on we will use the Planck units normalized by $L =1$. With this choice the energy density $\varrho$ and the pressure $p$ are
\begin{equation}\lbel{e:91}
	\varrho(t)=p(t)=-\frac{3}{4}R(t) = \frac{1}{2}\left[\frac{p_{\phi}(t)}{t}\right]^2,
\end{equation}
and the spacetime singularity occurs for either of $t\to0^{\pm}$. The positivity of $N$ implies that the future-pointing evolution of the scalar field is ``into'' a \tit{big crunch} for $t\in\mbb{R}^-$ and ``away from'' a \tit{big bang} for $t\in\mbb{R}^+$. From \eref{e:14} it is evident that for an element of the branch of the solution space admitting a big crunch the value of $\phi(t)-\phi^{\star}$ for $p_{\phi}^{\star}\in\mbb{R}^-$ is related to the analogous value for $p_{\phi}^{\star}\in\mbb{R}^+$ by an overall sign-change. The same holds for an element of the big bang branch of the solution space. Furthermore, we have the correspondence
\begin{equation}\lbel{e:1002}
	(\sig(p_{\phi}^{\star})[\phi(t)-\phi^{\star}])_-=-(\sig(p_{\phi}^{\star})[\phi(t)-\phi^{\star}])_+,
\end{equation}
which relates big crunches (the left-hand side for $t\in\mbb{R}^-$) with big bangs (the right-hand side for $t\in\mbb{R}^+$). In the canonical formalism at hand these identities are manifestations of the invariance of the \tit{covariant} action under a replacement of $\phi$ with $-\phi$.

Finally, we note again that the variables $t$ and $p_{\phi}$ are dimensionless but still not invariant under a $\zeta$-transformation. The observable scalar field in \eref{e:14} and the spacetime Ricci scalar in \eref{e:1003} -- along with its related scalars in \eref{e:91} -- are inheriting this implicit non-invariance. This fact will play a crucial 
role in singling out the correct regularization scheme in the quantum theory.

\section{\lbel{s:11}Quantum theory}

Our goal here is to build the precise quantum mechanical representation of the model introduced above. This means in particular the construction of a suitable Hilbert space $\msf{H}$ and the representation of the Hamiltonian $H_{\txt{S}}$ [see \eref{e:1001}] as a self-adjoint operator acting on a suitable domain in $\msf{H}$. The quantum evolution will then be determined by a Schr\"odinger equation
\begin{equation}\lbel{e:13}
	\ima\frac{\pde}{\pde t}\psi= \widehat{\left|\frac{p_{\phi}}{t}\right|}\psi=\hat{H}_{\txt{S}}\psi
\end{equation}
for $\psi\equi\psi(t,\phi)$.

\subsection{The scalar field momentum operator}

To begin, let us recall that the canonical formalism introduced in the previous section describes a freely evolving isotropic and flat FLRW model. The evolution is governed by the Hamiltonian $H_{\txt{S}}$, which by \eref{e:73} depends on the scale factor clock $t=V_{\txt{C}}a^3/|T^{\dms}|$. The scalar field $\phi$ is thus the only object subject to a quantization. Here, we have several possibilities to proceed.

The most obvious way to construct the quantum description is to apply the Schr\"odinger representation as in \cite{kreienbuehl:09}. As shown there, this representation leads neither to a singularity avoidance nor to a singularity resolution as the semiclassical wave packets simply follow the classical trajectories.

An alternative approach that is pursued here is the implementation of the polymer representation \cite{ashtekar:03a,halvorson:03}. As we will see, the requirement of the existence of an infrared regulator removal limit (see the previous section) forces this representation to be time-\tit{dependent}.\footnote{This situation is analogous to the one in the loop quantization of the geometric degrees of freedom, where consistency requirements label the improved dynamic construction as the correct one \cite{cs-con}.}

To initiate the detailed specification of the polymer quantization procedure, let us briefly recall the standard Schr\"odinger representation. It is characterized by the Stone-von Neumann uniqueness theorem \cite{reed:81,ashtekar:03a,halvorson:03}, which implies that among all the irreducible regular realizations of the Weyl form
\begin{equation}\lbel{e:2}
	\hat{I}_{\lambda}\hat{J}_{\mu}=e^{\ima\lambda\mu}\hat{J}_{\mu}\hat{I}_{\lambda},\quad\lambda,\mu\in\mbb{R}^+,
\end{equation}
of the canonical commutation relation
\begin{equation}\lbel{e:1}
	[\hat{\phi},\hat{p}_{\phi}]=\ima\hat{1}
\end{equation}
on the space $\msf{L}^2(\mbb{R},\txt{d}\phi)$ of Lebesgue square-integrable functions, the Schr\"odinger representation 
\begin{equation}\lbel{e:4}
	\hat{I}_{\lambda}\equi e^{\ima\lambda\hat{\phi}},\quad\hat{J}_{\mu}\equi e^{-\ima\mu\hat{p}_{\phi}},
\end{equation}
is unique up to unitary transformations. The regularity property says that the mappings of $\lambda$ to $\hat{I}_{\lambda}$ and $\mu$ to $\hat{J}_{\mu}$ are (strongly) continuous, which holds if for $\psi,\omega\in\msf{L}^2(\mbb{R},\txt{d}\phi)$ the mappings
\begin{equation}\lbel{e:5}
	\lambda\mapsto\langle\psi|\hat{I}_{\lambda}|\omega\rangle,\quad\mu\mapsto\langle\psi|\hat{J}_{\mu}|\omega\rangle,
\end{equation}
are (weakly) continuous. 

To generalize the above formalism, let us now consider the space of exponentiated operator labels, further parametrized in the following way
\begin{equation}\lbel{e:28}
  \lambda\equi\lambda_t\equi\nu\nu_t,\quad\mu\equi\mu_t\equi\rho\rho_t,\quad\nu,\nu_t,\rho,\rho_t\in\mbb{R}^+.
\end{equation}
The subscript $t$ can be seen as parametrizing the parameters $\lambda$ and $\mu$ of the groups of unitary operators $\hat{I}_{\lambda}$ and $\hat{J}_{\mu}$, respectively. \tit{That is, the Weyl algebra given in \eref{e:2} depends now on time and so do the unitary operators defined in \eref{e:4}.} However, because of the (strong) continuity of the operators $\hat{I}_{\lambda}$ and $\hat{J}_{\mu}$ in $\lambda$ and $\mu$, respectively, the operators $\hat{\phi}$ and $\hat{p}_{\phi}$ are
\begin{subequations}\lbel{e:25}
\begin{align}
	\hat{\phi}&\equi\ima\lim_{\lambda_t\to0}\frac{\hat{1}-\hat{I}_{\lambda_t}}{\lambda_t}=\ima\lim_{\nu\to0}\frac{\hat{1}-\hat{I}_{\lambda_t}}{\lambda_t}=\phi\hat{1},\\
	\hat{p}_{\phi}&\equi-\ima\lim_{\mu_t\to0}\frac{\hat{1}-\hat{J}_{\mu_t}}{\mu_t}=-\ima\lim_{\rho\to0}\frac{\hat{1}-\hat{J}_{\mu_t}}{\mu_t}=-\ima\frac{\pde}{\pde\phi},
\end{align}
\end{subequations}
thus they are independent of $t$. To conclude, in the Schr\"od\-ing\-er quantization the operators $\hat{\phi}$ and $\hat{p}_{\phi}$ are not changing if the group parameters are themselves parametrized by $t$. In particular, they remain time-independent.

The situation changes drastically if the regularity condition in the Stone-von Neumann uniqueness theorem is dropped. In this case a possible faithful realization of \eref{e:2} is given by the so-called ``polymer representation''. For the sake of generality we will further allow it to be time-\tit{dependent} (in a yet unspecified way as it was the case above). The \tit{non}-separable Hilbert space $\msf{H}$ for this representation consists of functions $\psi\in l^2(\mbb{R},\#_{\phi})$ satisfying the square-\tit{summation} requirement
\begin{equation}\lbel{e:3}
	\|\psi\|^2\equi\smashoperator[lr]{\sum_{\phi\in\mbb{D}_{\mu_t}(\psi)}}|\psi(\phi)|^2<\infty.
\end{equation}
The inner product on $\msf{H}$ providing this norm is
\begin{equation}\lbel{e:19}
	\langle\psi|\omega\rangle\equi\smashoperator[lr]{\sum_{\phi\in\mbb{D}_{\mu_t}(\psi,\omega)}}\overline{\psi(\phi)}\omega(\phi),
\end{equation}
where $\omega$ is another element of $l^2(\mbb{R},\#_{\phi})$. We denote by $\#_{\phi}$ the measure that maps a subset of $\mbb{R}$ to its cardinality (the so-called ``counting measure'') and by
\begin{equation}
	\mbb{D}_{\mu_t}(\psi)\equi\smashoperator[lr]{\bigcup_{\phi_0\in\mbb{S}(\psi)/\eqre}}\mbb{L}_{\mu_t}(\phi_0),\quad\mbb{D}_{\mu_t}(\psi,\omega)\equi\smashoperator[lr]{\bigcup_{\phi_0\in[\mbb{S}(\psi)\cap\,\mbb{S}(\omega)]/\eqre}}\mbb{L}_{\mu_t}(\phi_0),
\end{equation}
domains defined in terms of (necessarily countable) supports $\mbb{S}$ of $\psi$ and $\omega$. For $\phi,\chi\in\mbb{R}$ we define by $\phi\eqre\chi$ 
an equivalence relation such that $\phi$ and $\chi$ are equivalent if and only if there exists an integer $k\in\mbb{Z}$ such that $\phi=\chi+k\mu_t$ [recall \eref{e:28}]. The domains $\mbb{D}_{\mu_t}$ are then disjoint unions of uniform lattices\footnote{The role of these sets will become evident in the next subsection.}
\begin{equation}\lbel{e:22}
	\mbb{L}_{\mu_t}(\phi_0)\equi\{\phi_0\}+\mbb{Z}\mu_t,\quad\phi_0\in[0,\mu_t).
\end{equation}
Orthonormal basis states of $\msf{H}$ are ``half-deltas''
\begin{equation}\lbel{e:77}
	\delta_{\phi}:\chi\mapsto\delta_{\phi}(\chi)\equi\delta_{\phi\chi}\equi\begin{cases}
		1, & \phi=\chi,\\
		0, & \txt{otherwise},
	\end{cases}
\end{equation}
extending the definition of the Kronecker delta symbol to the real line. 

The polymer representation is now given by
\begin{equation}\lbel{e:6}
	\hat{I}_{\lambda_t}\delta_{\phi}\equi e^{\ima\lambda_t\phi}\delta_{\phi},\quad\hat{J}_{\mu_t}\delta_{\phi}\equi\delta_{\phi+\mu_t},
\end{equation}
which characterizes again a \tit{multiplication} and a \tit{translation} operator, respectively. The ``$\lambda$-mapping'' given in \eref{e:5} is once more continuous in $\lambda_t$ so that by Stone's theorem \cite{reed:81,halvorson:03} the scalar field multiplication operator remains to be given by \eref{e:25}. The difference to the Schr\"odinger representation in \eref{e:4} is that the ``$\mu$-mapping'' in \eref{e:5} is no longer continuous in $\mu_t$. There is therefore \tit{no} self-adjoint momentum operator generating infinitesimal translations. On $l^2(\mbb{R},\#_{\phi})$ there is only an operator generating \tit{finite} translations. We are therefore forced to regularize it, for which we employ the technique introduced by Thiemann in the context of full LQG \cite{t-afree,*t-qsd5,thiemann:08}. In essence this technique is approximating the undefined $\hat{p}_{\phi}$ by well-defined translation operators. Following \cite{ashtekar:03a,corichi:07,hossain:10b}, we choose
\begin{subequations}\lbel{e:1234567}
\begin{align}
	\hat{p}_{\phi\mu_t}&\equi-\frac{\ima}{2\mu_t}(\hat{J}_{\mu_t}^{\dag}-\hat{J}_{\mu_t}),\\
	\widehat{p_{\phi\mu_t}^2}&\equi\frac{2}{\mu_t^2}\left(\hat{1}-\frac{\hat{J}_{\mu_t}^{\dag}+\hat{J}_{\mu_t}}{2}\right).
\end{align}
\end{subequations}
The action of the former on a state $\psi\in\msf{H}$ is
\begin{equation}\lbel{e:26}
	\hat{p}_{\phi\mu_t}\psi(\phi)=-\frac{\ima}{2\mu_t}[\psi(\phi+\mu_t)-\psi(\phi-\mu_t)]
\end{equation}
so that, \tit{if} we could send $\mu_t$ to $0$ [or according to \eref{e:28} send $\rho$ to $0$, thereby taking the limit at the \tit{kinematic} level], we would get back the differential operator $-\ima\pde/(\pde\phi)$. We observe that the representation of the momentum operator $\hat{p}_{\phi\mu_t}$ is highly non-unique, in the same way the representation of finite difference operators in numerical analysis is. We stress that, unlike in the Schr\"odinger representation, the momentum operator is now time-\tit{dependent} [see \eref{e:28}].

At this point it is necessary to emphasize that the presented polymer quantization is not the only possible one. Essentially, by replacing the roles of $\phi$ and $p_{\phi}$ we arrive at another polymer representation, \tit{inequivalent} (and in a sense ``dual'') to ours (see the discussion in \cite{halvorson:03}). Such a dual representation was used in the quantization of the scalar field in full LQG \cite{klb-scalar,*klo-scalar}. Its symmetry reduced version was applied to the LQC model of an FLRW universe \cite{ddl-scalar} filled with a massless scalar field. The subsequent analysis of the spectral decomposition of the evolution operator (playing the role of the Hamiltonian) shows that the dynamic of such a system is \tit{exactly the same as} the one of the system with the scalar field quantized via standard methods of quantum mechanics \cite{ashtekar:06c}. Both approaches, ours and the one of \cite{ddl-scalar}, are equally viable from a mathematical point of view. Therefore, choosing one of them requires a physical input.

\subsection{The Hamiltonian}

The next step is the construction of the quantum Hamiltonian $\hat{H}_{\txt{S}}$ and the determination of its action, which generates the unitary evolution. To do so, we switch to the scalar field momentum space, which is again the Pontryagin dual of the real line but this time the latter is equipped with the discrete topology. In short, it is the Bohr-compactified real line $\mbb{R}_{\txt{B}}$. The Hilbert space defined in the previous subsection is then equivalent to the space $\msf{H}^{\nal}$, which consists of Bohr square-measurable functions
\begin{equation}
	\psi^{\nal}(p_{\phi})\equi\smashoperator[lr]{\sum_{\phi\in\mbb{D}_{\mu_t}(\psi)}}\psi(\phi)e^{-\ima\phi p_{\phi}}\in\msf{L}^2(\mbb{R}_{\txt{B}},(\txt{d}p_{\phi})_{\txt{B}})
\end{equation}
satisfying
\begin{align}\lbel{e:63}
	\|\psi^{\nal}\|^2&\equi\int|\psi^{\nal}(p_{\phi})|^2\;(\txt{d}p_{\phi})_{\txt{B}}\nonumber\\
	&\equi\lim_{C\to\infty}\frac{1}{2C}\int_{-C}^C|\psi^{\nal}(p_{\phi})|^2\;\txt{d}p_{\phi}<\infty.
\end{align}
The inner product (between $\psi^{\nal}$ and another $\omega^{\nal}\in\msf{H}^{\nal}$) generating this norm is
\begin{equation}\lbel{e:20}
	\langle\psi^{\nal}|\omega^{\nal}\rangle\equi\int\overline{\psi^{\nal}(p_{\phi})}\omega^{\nal}(p_{\phi})\;(\txt{d}p_{\phi})_{\txt{B}}.
\end{equation}
The basis orthonormal with respect to it is formed by the plane waves
\begin{equation}
	e_{\phi}:p_{\phi}\mapsto e_{\phi}(p_{\phi})\equi e^{-\ima\phi p_{\phi}}=\delta_{\phi}^{\nal}(p_{\phi}).
\end{equation}
We observe that for a uniform lattice $\mbb{D}_{\mu_t}(\psi)=\mbb{L}_{\mu_t}(\phi_0)$ [see \eref{e:22}] the Bohr measure $(\txt{d}p_{\phi})_{\txt{B}}$ becomes the Lebesgue measure $\txt{d}p_{\phi}$ with an integration over the fixed interval $(-\pi/\mu_t,\pi/\mu_t]$. The momentum space polymer theory defined here \tit{would} then be that of Fourier with discreteness in position rather than momentum space.

In the general polymer theory at hand, the action of the multiplication and translation operator on the basis states $e_{\phi}$ is unchanged in comparison to \eref{e:6} so that the scalar field operator is given by $\ima\pde/(\pde p_{\phi})$. On the other hand, the (undefined) operator $\hat{p}_{\phi}$ has become the regularized $\hat{p}_{\phi\mu_t}$ (see the previous subsection), which is approximated by translation operators according to \eref{e:1234567}. Since $e_{\phi+\mu_t}=e_{\mu_t}e_{\phi}$ we obtain
\begin{equation}\lbel{e:71}
	\hat{J}_{\mu_t}=e^{-\ima\mu_tp_{\phi}}\hat{1},\quad\hat{p}_{\phi\mu_t}=\frac{\sin(\mu_tp_{\phi})}{\mu_t}\hat{1},
\end{equation}
because of which the action of the Hamiltonian operator can be explicitly given by
\begin{equation}\lbel{e:17}
	\hat{H}_{\txt{S}}\psi^{\nal}=\left|\frac{\sin(\mu_tp_{\phi})}{\mu_tt}\right|\psi^{\nal},
\end{equation}
where $\psi^{\nal}\equi\psi^{\nal}(t,p_{\phi})$. The fact that $\|\psi\|=\|\psi^{\nal}\|$ allows us now to prove the conservation of the norm under an action of $\hat{H}_{\txt{S}}$. To show this we write explicitly the time derivative of the norm
\begin{align}\lbel{e:60}
	&\ima\frac{\pde}{\pde t}\|\psi\|^2=\ima\frac{\pde}{\pde t}\|\psi^{\nal}\|^2=\int\ima\frac{\pde}{\pde t}|\psi^{\nal}|^2\;(\txt{d}p_{\phi})_{\txt{B}}\nonumber\\
	&\quad=\smashoperator[lr]{\sum_{\phi,\chi\in\mbb{D}_{\mu_t}(\psi)}}\overline{\psi(t,\phi)}\psi(t,\chi)\int\hat{H}_{\txt{S}}e^{-\ima p_{\phi}(\phi-\chi)}\;(\txt{d}p_{\phi})_{\txt{B}}.
\end{align}
To evaluate the right-hand side we first observe that the integral can be expressed as the integral
\begin{equation}\lbel{e:61}\begin{split}
	&\lim_{C\to\infty}\frac{1}{2C}\int\limits_{-C}^C=\lim_{n\to\infty}\frac{\mu_t}{4n\pi}\sum_{k=1}^{n}\Bigg(\int\limits_{-2k\pi/\mu_t}^{-(2k-1)\pi/\mu_t} \\
	&\quad+\int\limits_{-(2k-1)\pi/\mu_t}^{-2(k-1)\pi/\mu_t}+\int\limits_{2(k-1)\pi/\mu_t}^{(2k-1)\pi/\mu_t}+\int\limits_{(2k-1)\pi/\mu_t}^{2k\pi/\mu_t}\Bigg).
\end{split}\end{equation}
The specific form of this integral allows us to drop the absolute value in \eref{e:17}, replacing it in \eref{e:60} instead with a sign appropriate for each integration domain in \eref{e:61}. Next, we apply some trigonometric identities, the $\mu_t$-translation invariance of $\mbb{D}_{\mu_t}(\psi)$, and (see \cite{gradshteyn:07})
\begin{equation}\lbel{e:62}
	\sum_{k=1}^{n}\cos\left(\frac{(2k+1)\pi}{\mu_t}(\phi-\chi)\right)=\frac{\sin\left(\cfrac{2n\pi}{\mu_t}(\phi-\chi)\right)}{2\sin\left(\cfrac{\pi}{\mu_t}(\phi-\chi)\right)}.
\end{equation}
Finally, if we divide this by $n$ and take the limit $n\to\infty$ [see \eref{e:61}], we get $\ima\pde\|\psi\|/\pde t=0$.

Up to now, the time-dependent shift parameter $\mu_t$ in the approximated $\hat{p}_{\phi}$ operator has been arbitrary. At the mathematical level the situation is analogous to the one in the loop quantization of the geometry (see \cite{abl-lqc}), where the fiducial holonomy length could be an arbitrary function on the phase space. There, however, the physical consistency requirements restricted the possible choices to just one class of functions \cite{cs-con}. We expect that the same situation occurs in our model. To show that this expectation is indeed realized let us recall the following facts.

The particular moment of the universe's evolution can be represented by various points on the phase space corresponding to different choices of the regulator cell. Furthermore, once we ask about the locally measurable properties of the universe (observables) at this moment, there has to exist their nontrivial limit as we remove the regulator.

One such local observable is the energy density \eref{e:91} determined by \eref{e:1003}. According to \eref{e:1001} and \eref{e:91}, the quantum operator corresponding to it is related to the Hamiltonian $\hat{H}_{\txt{S}}$ in the following way
\begin{equation}
   \hat{\varrho} = \frac{1}{2}\widehat{H_{\txt{S}}^2} .
\end{equation}
From \eref{e:17} it follows that at a fixed point in time $t$ the spectrum of this operator equals
\begin{equation}\lbel{e:sprho}
  {\rm Sp}(\hat{\varrho}) = \left[ 0 , \frac{1}{2\mu_t^2 t^2}\right] \ \subset\ \mbb{R}.
\end{equation}
The most natural way to satisfy the consistency requirements discussed in the previous paragraph is to require that $\txt{Sp}(\hat{\varrho})$ be time-\emph{independent}. This implies $\mu_t\propto1/t$ so that we can fix the function $\rho_t$ in \eref{e:28} by
\begin{equation}\lbel{e:15}
	\rho_t\equi\frac{1}{|t|}.
\end{equation}
This in turn gives $\mu_t=\rho/|t|$ [see again \eref{e:28}], which for the ``volume clock'' $v\equi t/\rho$ results in the momentum space Schr\"odinger equation
\begin{equation}\lbel{e:16}
	\ima\frac{\pde}{\pde v}\psi^{\nal}=\left|\sin\left(\frac{p_{\phi}}{v}\right)\right|\psi^{\nal}
\end{equation}
for $\psi^{\nal}\equi\psi^{\nal}(v,p_{\phi})$. 

Note that this method of fixing $\mu_t$ is almost a full analog of the conditions used for the geometry degrees of freedom in \cite{cs-con}. There, however, the reasoning exploited the existence of ``nicely'' behaving semiclassical sectors through the use of the so called \emph{effective dynamic}. Here, as we have not yet investigated the dynamical sector, implementing that reasoning directly would be risky. Instead, we managed to fix $\mu_t$ through considerations of the genuine quantum formalism.

In the next section we solve the Schr\"odinger equation \eref{e:16} in order to analyze the dynamic and to discuss the physical properties of the system.

\section{\lbel{s:0}The dynamic}

The Hilbert space and the explicit action of the Hamiltonian operator constructed just now allow us to determine the system's dynamic. At this level the requirement of the theory to be physically meaningful becomes crucial. The principal requirement is that the theory must have the proper low energy limit. Here, this means that in the distant past and future the quantum evolution ought to agree with the predictions of GR. In our case an inability of the model-description to realize this property would imply that the formulation should be further and adequately corrected. In fact, as we will see below, this is precisely what is required here.

To begin, let us investigate the dynamic of the theory exactly as specified in the previous section.

\subsection{Single lattice Hilbert space}

Once we select $p_{\phi}$ as the configuration variable, the Schr\"odinger equation given in \eref{e:16} becomes an ordinary differential equation, which we can solve for $v\in\mbb{R}^{\pm}$. The solution reads
\begin{subequations}\lbel{e:75}
\begin{align}\lbel{e:uni-form}
  \begin{split}
	\psi^{\nal}(v,p_{\phi})&\equi\hat{E}_{vv^{\star}}\psi^{\nal}(v^{\star},p_{\phi})\\
	&\equi e^{-\ima[F(v,p_{\phi})-F(v^{\star},p_{\phi})]}\psi^{\nal}(v^{\star},p_{\phi}),
  \end{split}\\
	F(v,p_{\phi})&\equi vS(v,p_{\phi})\left[\sin\left(\frac{p_{\phi}}{v}\right)-\txt{Ci}\left(\left|\frac{p_{\phi}}{v}\right|\right)\frac{p_{\phi}}{v}\right],\\
	S(v,p_{\phi})&\equi\sig\left(\sin\left(\frac{p_{\phi}}{v}\right)\right),
\end{align}
\end{subequations}
where we set $v^{\star}=t^{\star}/\rho$. For $|\txt{arg}(z)|<\pi$ the cosine integral function is
\begin{equation}\lbel{e:0000}
	\txt{Ci}(z)\equi\gamma+\ln(z)+\int_0^z\frac{\cos(y)-1}{y}\;\txt{d}y
\end{equation}
with $\gamma$ being the Euler-Mascheroni number \cite{abramowitz:72}. This definition implies $\txt{Ci}(z)\sim\gamma+\ln(z)$ for $z\to0$, suggesting semiclassical behavior of sufficiently sharply peaked initial states in the limit $|v|\to\infty$. 

The operator $\hat{H}_{\txt{S}}$ is not defined at $v=0$ and therefore neither is the ordinary differential equation \eref{e:16}. However, we have
\begin{equation}\lbel{e:102}
	\lim_{v\to0^{\pm}}F(v,p_{\phi})=0
\end{equation}
so that the solutions to \eref{e:16} [given in \eref{e:75} for $v\neq0$] have a ``removable singularity'' at $v=0$ \cite[Theorem 10.20]{rudin:1987}. More generally, this can be explained by means of Carath\'eodory's existence theorem \cite[Theorem 7.18]{rudin:1987}.

From \eref{e:102} it follows that we can define $F(0,p_{\phi})\equi0$, which implies that \tit{there exists a unique unitary operator}
\begin{equation}\lbel{e:74}
	\hat{E}_{0v^{\star}}\equi e^{\ima F(v^{\star},p_{\phi})}\hat{1}
\end{equation}
\tit{that evolves states to the instant $v=0$. In consequence,} \emph{there exists a preferred extension of the evolution through $v=0$}, defined by the requirement of continuity of $\psi^{\nal}$ at $v=0$. The global solution is thus given by \eref{e:75} with \eref{e:74}.

It appears that the existence of such a preferred extension is sufficient for singularity resolution. However, as we will see below, this is not the case. To explain what is missing, we consider any unit-normalized initial state $\psi^{\nal}(v^{\star},p_{\phi})$ such that the expectation value of the scalar field operator is finite
\begin{equation}
  \langle\psi^{\nal},v^{\star}|\hat{\phi}|\psi^{\nal},v^{\star}\rangle = \phi^{\star} .
\end{equation}
The expectation value of $\hat{\phi}$ at any value of $v$ is then given by the formula
\begin{align}\lbel{e:phi-evo}
	\langle\hat{\phi}\rangle_{\psi^{\nal}}(v) &\equi \langle\psi^{\nal},v|\hat{\phi}|\psi^{\nal},v\rangle=\left\langle\psi^{\nal},v^{\star}\left|\hat{\phi}-\left[S(v,p_{\phi})\txt{Ci}\left(\left|\frac{p_{\phi}}{v}\right|\right)\right.\right.\right.\nonumber\\
	&\quad\left.\left.\left.-S(v^{\star},p_{\phi})\txt{Ci}\left(\left|\frac{p_{\phi}}{v^{\star}}\right|\right)\right]\hat{1}\right|\psi^{\nal},v^{\star}\right\rangle.
\end{align}
Since the cosine integral function defined in \eref{e:0000} belongs to $\msf{L}^2(\mbb{R},\txt{d}p_{\phi})$, this expectation value is in fact equal to $\phi^{\star}$. That is to say the evolution is \tit{frozen}. This result is then in direct disagreement with the predictions of GR. In consequence, our states exhibit an unphysical behavior in the low energy (large $|v|$) limit.

Our model then still lacks an appropriate \tit{physical} Hilbert space. 
To explore the possibilities of constructing it, let us first go back to analyzing the 
solutions to \eref{e:16} but this time by considering the wave functions on the configuration 
space as opposed to the momentum one used in \eref{e:75}. On the configuration space, the evolution of a state $\psi_v\equi\psi(v,\cdot)$ can be viewed as an assignment $v\mapsto\psi_v\in\msf{H}_v$, where $v\in\mbb{R}$. The Hilbert space $\msf{H}_v$ is spanned by eigenstates of $\hat{H}_{\txt{S}}$ for a fixed value of $v$. However, per analogy with the loop quantization of the geometry \cite{ashtekar:06c} we can distinguish sectors that are invariant with respect to the action of $\hat{H}_{\txt{S}}$ at $v$. These sectors consist of functions that are supported on the lattices $\mbb{L}(\varphi_0)\mu_v$, where $\mu_v=1/|v|$ [see \eref{e:28}, \eref{e:15}, and the definition of $v$ in the sentence prior to \eref{e:16}] and
\begin{equation}
	\mbb{L}(\varphi_0)\equi\mbb{L}_{\mu_v}(\phi_0)/\mu_v\equi\{\varphi_0\}+\mbb{Z},\quad\varphi_0\in[0,1).
\end{equation}
We can then regard at the initial $v=v^{\star}$ the subspaces $\msf{H}_{v\varphi_0} \equi\msf{H}_v|_{\mu_v\mbb{L}(\varphi_0)}$ as the superselection sectors and evolve them independently. Such a decomposition 
can be performed at each $v$ independently.
Let us now probe whether there exists any relation between the spaces $\msf{H}_{v\varphi_0}$ for different values of $v$. The answer is given by the form of \eref{e:75}: since the cosine integral function is non-periodic, the unitary evolution to any $v^{\star}+v_{\varepsilon}$, where $v_{\varepsilon}\in\mbb{R}^{\pm}$, instantaneously couples an infinite number of these lattices. In consequence, the sectors $\msf{H}_{v\varphi_0}$ of $\msf{H}_v$ are \tit{not} true superselection sectors in the sense of \cite{ashtekar:06b,ashtekar:06a,ashtekar:06c,ashtekar:07,ashtekar:11}. Therefore, we are forced to work with the original \emph{non-separable} Hilbert space $\msf{H}_v$ without access to previously available tools that allow for a distinction of separable subspaces. The form of \eref{e:phi-evo} suggests then that in order to provide a nontrivial evolution, the physical Hilbert space needs to be equipped with a continuous rather than a discrete inner product.

\subsection{\lbel{s:1000} Integral Hilbert space}

A similar situation appeared in LQC already in a different context during the studies of the FLRW universe with a massless scalar field and a positive cosmological constant \cite{pa-posL,kp-posL}. There, following the choice of a lapse adopted to using the scalar field as time variable, the evolution operator admitted a family of self-adjoint extensions, each with a discrete spectrum. However, a different choice of the lapse -- corresponding to parametrizing the evolution by the cosmic time variable -- led to a unique self-adjoint generator of the evolution with a continuous spectrum \cite{kaminski:09}. The physical Hilbert space corresponding to the latter case (the ``cosmic time case'') appeared, furthermore, to be an integral of all the Hilbert spaces corresponding to the particular self-adjoint extensions of the former case (the ``matter clock case''), with the Lebesgue measure determined by the group averaging procedure. 

Motivated by this observation, we introduce the analog of the integral Hilbert space in our case.
First, we note that on the domain $[0,1)$ of $\varphi_0$ one can introduce a natural (quite general and time dependent) Lebesgue measure $M(v,\varphi_0)\txt{d}\varphi_0$. 
Next, we introduce a decomposition of the non-separable Hilbert space $\msf{H}$ into spaces $\msf{H}_{v\varphi_0}$ at the initial time $v^{\star}$. This leads to the following definition of the decomposition of the initial data at $v=v^{\star}$
\begin{equation}
  \msf{H}_{v^{\star}}\ni\psi(v^{\star},\phi) \mapsto \psi_{\varphi_0}(v^{\star},\phi)\equi\psi(v^{\star},\phi)|_{\mbb{L}_{\mu_{v^{\star}}}(\phi_0)}
  \in \msf{H}_{v^{\star}\varphi_0} .
\end{equation}
This initial data is then extended to the solutions to \eref{e:16} via \eref{e:75}. We thus have a decomposition of the physical Hilbert space into explicitly separable (at least at $v=v^{\star}$) subspaces.

Now, we can define the new \tit{physical} Hilbert space $\msf{H}_{\txt{P}v}$ at $v=v^{\star}$ via
\begin{equation}
	\msf{H}_{\txt{P}v^{\star}}\equi\int_0^1\msf{H}_{v^{\star}\varphi_0}M(v^{\star},\varphi_0)\;\txt{d}\varphi_0 
\end{equation}
and equip it with the inner product
\begin{equation}
	\langle\psi_v|\omega_v\rangle\equi\int_0^1\langle\psi_{v\varphi_0}\hat{E}_{vv^\star}|\hat{E}_{v^\star v}\omega_{v\varphi_0}\rangle M(v^{\star},\varphi_0)\;\txt{d}\varphi_0.
\end{equation}
We used the abbreviation $\psi_{v\varphi_0} \equi \psi_{\varphi_0}(v,\phi) \in \msf{H}_{v\varphi_0}$.
This is our candidate for the physical inner product: between each pair of solutions it is evaluated on the initial data slice at $v=v^{\star}$. On that initial slice it can be written as
\begin{equation}\lbel{e:phip}
	\langle\psi_{v^{\star}}|\omega_{v^{\star}}\rangle\equi\int_{\mbb{R}}\overline{\psi(v^{\star},\phi)}\omega(v^{\star},\phi)M(v^{\star},\varphi_0(\phi))/\mu_{v^{\star}}\;\txt{d}\phi.
\end{equation}
It is by definition time-independent but a priori it may not have a local form analogous to \eref{e:phip} at $v\neq v^{\star}$, which can potentially complicate the evaluations of the expectation values of the observables.

We note that the construction performed for $v=v^{\star}$ can be repeated at each value 
of $v$, giving rise to potentially inequivalent constructions of the candidate physical Hilbert space. One can then consider a function
\begin{equation}
	P(\psi_v|\omega_v) \equi \int_{\mbb{R}}\overline{\psi(v,\phi)}\omega(v,\phi)M(v,\varphi_0(\phi))/\mu_v\;\txt{d}\phi.
\end{equation}
On each slice of constant $v$ this function equals the inner product of the candidate Hilbert space constructed for this slice. One can then ask under which condition these Hilbert spaces will be equivalent and their inner products equal. A condition necessary and sufficient for it
is that $\partial P(\psi_v|\omega_v)/\pde v = 0$. 
The form of the unitary evolution operator [see in particular \eref{e:uni-form}] implies however that this condition will be satisfied if and only if we require
\begin{equation}
  M(v,\varphi_0)\equi\mu_vm(\varphi_0) . 
\end{equation}
Following this choice, our candidate Hilbert space becomes (up to a rescaling $\psi_v\mapsto\psi_v/m^{1/2}$ on $\msf{H}_v$) the space $\msf{L}^2(\mbb{R},\txt{d}\phi)$ with the standard $\msf{L}^2$-inner product. Also, the \tit{momentum} space is now $\msf{L}^2(\mbb{R},\txt{d}p_{\phi})$ with the corresponding Lebesgue measure.

Using \eref{e:74}, we can now consider an initial state
\begin{equation}\lbel{e:4141}
	\psi^{\nal}(v^{\star},p_{\phi})\equi\hat{E}_{v^{\star}0}\psi^{\nal}(0,p_{\phi})
\end{equation}
with a \tit{real} unit-$\msf{L}^2$-normalized Gaussian
\begin{equation}\lbel{e:141414}
	\psi^{\nal}(0,p_{\phi})\equi\sqrt{\frac{w}{\sqrt{\pi}}}e^{-w^2(p_{\phi}-p_{\phi}^{\star})^2/2}.
\end{equation}
This class of states is ``special'' in the sense that the quantum evolution they undergo is semiclassical both for $v\sim v^{\star}$ and $v\sim-v^{\star}$ (see below). Furthermore, the states $\hat{E}_{v0}\psi^{\nal}(0,p_{\phi})$ with unit-$\msf{L}^2$-normalized $\psi^{\nal}(0,p_{\phi})$ span the solution space of the Wheeler-DeWitt equation defined by the Hamiltonian constraint in \eref{e:1004} for the clock $V_{\txt{C}}a^3=T$. The set of the complex conjugate of these states represents the analogous states for the clock $-V_{\txt{C}}a^3=T$ (see \cite{catren:06} and recall that we set $L =1$). They are thus particularly convenient in comparing the evolution in the Schr\"odinger and polymer quantizations.

\begin{figure}
\begin{center}
	\subfigure[\lbel{f:01} Quantum evolution for two values of $p_{\phi}^{\star}$.]{\includegraphics[scale=1]{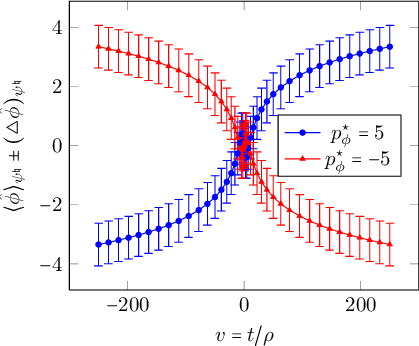}}
	\subfigure[\lbel{f:02} Classical and quantum evolution for $p_{\phi}^{\star}=5$.]{\includegraphics[scale=1]{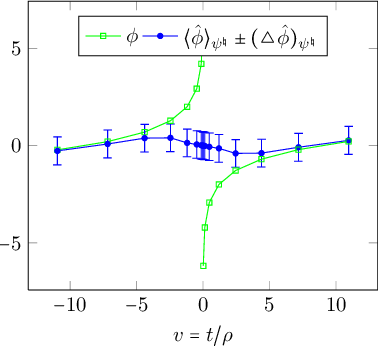}}
	\caption{\lbel{f:0}An illustration of the quantum evolution of the scalar field is presented in Fig.\;\ref{f:01}. In Fig.\;\ref{f:02} we display the $p_{\phi}^{\star}=5$ quantum evolution along with the corresponding classical evolution over a subset of the $v$-interval used in Fig.\;\ref{f:01}. The circles, triangles, and squares correspond to actual measurements made using the software {\scshape Matlab} and GNU Octave. The time interval between two consecutive measurements becomes smaller as $|v|$ approaches $0$. In this plot, the smallest value of $|v|$ is $0.25$ but values as small as $10^{-4}$ have been considered with the same outcome. Namely, there is semiclassicality for $|v|\gg1$ and the big bang singularity at $v=0$ is resolved.}
\end{center}
\end{figure}

We can now try to evaluate once more the expectation value of the scalar field. For that we choose for \eref{e:4141} the initial ``time'' $v^{\star}=250$ and for \eref{e:141414} the width $w=1$. This results in a relatively small initial value of both the scalar field and the momentum fluctuations (see below). We further set $p_{\phi}^{\star}=5$ to prevent any significant portion of the Gaussian initial state from overlapping with the momentum space origin, as this is where the cosine integral function has an integrable singularity. 
The quantum evolution of such an initial state can then be calculated numerically. Figure \ref{f:0} shows the quantum trajectory corresponding to this evolution. From there, it is evident that the evolution is semiclassical for $|v|\gg1$. In fact, the classical solution that is well approximating the quantum trajectory is characterized by
\begin{equation}
	\phi^{\star}=\mp\left\langle S(v^{\star},p_{\phi})\left[\txt{Ci}\left(\left|\frac{p_{\phi}}{v^{\star}}\right|\right)+\ln(|v^{\star}|)\right]\hat{1}\right\rangle_{\psi^{\nal}},
\end{equation}
where the overall sign ``$\mp$'' corresponds to $v\in\mbb{R}^{\pm}$ and, where $v^{\star}=250$. Given the definition of the function $S$ in \eref{e:75}, we observe that $|v^{\star}|\gg|p_{\phi}^{\star}|/\pi$ is a necessary requirement for semiclassicality. 
As we can see, the physical state is indeed passing in a continuous manner \tit{through} the point $v=0$, corresponding in the classical theory to the big bang singularity [which is particularly clear from Fig.\;\ref{f:01} and also from \eref{e:102}].
This happens regardless of the sign of the initial momentum so that the quantum evolution is effectively respecting \eref{e:1002}, which in the classical theory specifies the relation between the solutions for negative and positive $t$.

\begin{figure}
\begin{center}
	\includegraphics{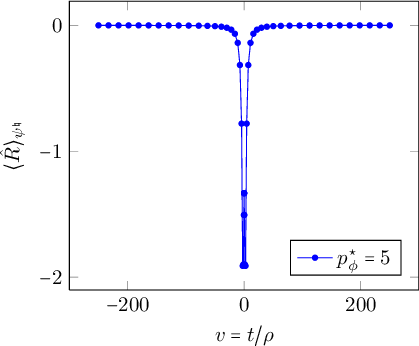}
	\caption{\lbel{f:1}Expectation values of the spacetime Ricci operator for various values of $v$ are plotted in this figure. As in Fig.\;\ref{f:0}, measurements are taken the more often the closer $|v|$ is to $0$ with the smallest values given by $|v|=0.25$. It follows that the big bang curvature singularity is resolved and the branches $v\in\mbb{R}^{\pm}$ are connected.}
\end{center}
\end{figure}

To examine more closely the issue of the singularity resolution we also analyzed the expectation values of the operator corresponding to the spacetime Ricci scalar. The quantum trajectory is presented in Fig.\;\ref{f:1}. The measurements are independent of the overall sign of $p_{\phi}^{\star}$, thus only the case $p_{\phi}^{\star}>0$ has been plotted. We see that the spacetime curvature remains finite for all $v$. This confirms the analytical result of \eref{e:sprho} and, thus, implies the global boundedness of the spectrum of the Ricci scalar operator once $\mu_v$ is fixed via \eref{e:15}. One can thus conclude that the big bang singularity is resolved.

\begin{figure}
\begin{center}
	\includegraphics{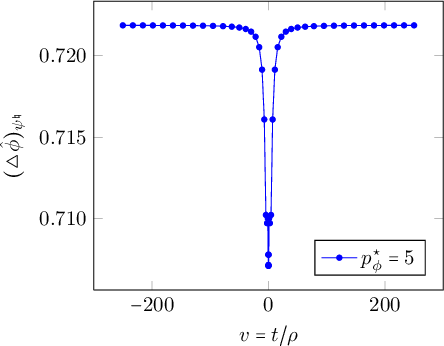}
	\caption{\lbel{f:2}In this figure an illustration of the fluctuations of the scalar field operator is presented. For $|v|\to\infty$ the fluctuations increase but remain nonetheless finite.}
\end{center}
\end{figure}

\begin{figure}
\begin{center}
	\includegraphics{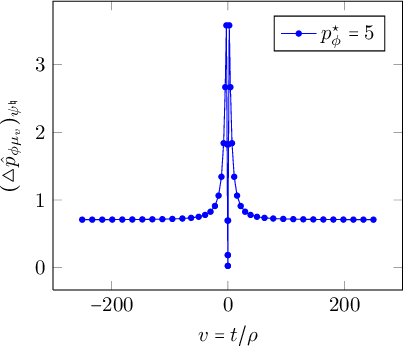}
	\caption{\lbel{f:3}This figure provides an illustration of the fluctuations of the polymeric momentum operator. Just like for the scalar field fluctuations in Fig.\;\ref{f:2}, the behavior for $|v|\gg1$ indicates semiclassicality.}
\end{center}
\end{figure}

Finally, in Figs.\;\ref{f:2} and \ref{f:3} the fluctuations of the scalar field and polymer momentum operator are depicted, respectively. They both quickly approach $(1/2)^{1/2}$ as $|v|$ increases, which is the value expected for a Gaussian with $w=1$. This confirms the semiclassical nature of the state for $|v|\sim v^{\star}$. What is interesting in the near-singularity region is the fact that the fluctuations of the scalar field operator are in fact decreasing for $|v|\to 0$. This may happen because the state gets ``squeezed'' towards the origin in order to ``fit through'' the point $v=0$. This however requires a more detailed analysis of the nature of the state there, which may be the subject of a subsequent investigation.

\section{\lbel{s:12}Discussion}

In this paper we investigated the quantum dynamic of the isotropic and flat FLRW universe of infinite extent and sourced by a minimally coupled massless scalar field. Our focus was on the modifications to the dynamic following from the polymeric nature of the matter, and in particular on the issue of the singularity resolution. To identify these effects we implemented one of two possible loop quantization schemes of the scalar field. This scheme is the analog of the one used so far in LQC to quantize the geometry degrees of freedom. \tit{Unlike} in the most of the existing works in LQC, instead of implementing the Dirac program to solve the Hamiltonian constraint, we performed a complete deparametrization of the system by choosing a time variable that depends on the scale factor. As a result, the physical evolution is described by a free Hamiltonian. The quantization of such a deparametrized system is implicitly equivalent to selecting the Schr\"odinger quantization for the geometry when applying the Dirac program. Therefore, the effects of the geometry discreteness are not featured in our model. Rather, the matter degrees of freedom are discreet.

In the process of constructing the correct description of the quantum system we encountered several obstacles:

First, the noncompactness of the universe's spatial slices forced us to introduce an infrared regulator. The necessary consistency condition, that the theory has to admit a well defined and nontrivial regulator removal limit, restricted then the Hamiltonian to a particular form, which happened to be explicitly time-dependent.  

Second, the Hilbert space to which the physical states belong occurred to be non-separable. This is a standard (and treatable) problem in LQC. Here, however, the explicit time dependence of the Hamiltonian prevented us from implementing the known technique of subdividing the (too big) Hilbert space onto separable superselection sectors. An idea to naively proceed by determining the dynamic on that space led to a model significantly disagreeing with GR predictions at the low curvature limit. Indeed, the quantum evolution of the scalar field \tit{was} frozen.

To cure this defect we performed a specific construction of the separable Hilbert space out of the nonseparable one, taking as the guideline the relation between Hilbert spaces corresponding to the models with different choices of the lapse function in LQC in the presence of a positive cosmological constant. As a result, we were able to construct a certain integral Hilbert space equipped with a continuous rather than a discrete (as usual in LQC) inner product. 

Such a construction of the Hilbert space was then used to investigate the dynamic. To do so we selected a class of Gaussian initial states and evolved them numerically. The resulting quantum trajectories showed a good convergence to the classical trajectories predicted by GR at low energies. At high curvatures (small $|v|$) however we observed a significant departure from GR. Indeed, the most critical feature of the model is the existence of a unique unitary evolution operator evolving to/from the time slice $v=0$ corresponding to the classical singularity. This and the regularity of the wave function describing the physical state allowed us to select a naturally preferred extension of the evolution, thus ensuring a \emph{deterministic evolution through the classical singularity.} Furthermore, the quantum counterparts of the Ricci scalar, energy density, or pressure are explicitly bounded operators. In consequence, the listed quantities remain finite throughout the entire evolution, including in particular $v=0$.

At this moment it is important to note that, \tit{unlike in previous contributions to the literature on this model, here the quantum features responsible for singularity resolution originate from the matter rather than the geometric sector.} Therefore, the form of the singularity resolution differs from that in the literature: instead of being avoided, the surface $v=0$ is made passable and all the standard locally measurable quantities remain finite. One has to remember, however, that the presented picture of high energy behavior is incomplete. Getting a more robust description requires taking into account the polymer nature of \emph{both} the matter and the geometry, in which situation the results may change qualitatively. The analysis of this scenario is a task for the future.

Finally, let us comment on an important lesson learned from this model: the predicted dynamic depends critically on the construction of the physical Hilbert space of the model, even though the regularized form of the Hamiltonian remains the same. This implies in particular that the regularized form of the classical Hamiltonian or Hamiltonian constraint \emph{is not sufficient} to robustly determine or even well approximate the quantum evolution. The problem in Hilbert space construction appears not only when the matter degrees of freedom are quantized ``\tit{\`a la} loop'' but already in ``standard'' LQC in the models as simple as a Bianchi I universe \cite{hp-b1}. This issue is particularly critical in all the studies of models in LQC performed via the so called \emph{effective dynamic} techniques without prior specification of the elements of the genuine quantum theory that the effective formulation is supposed to mimic.

In a further project we intend to take a closer look at the behavior of the state near the singularity. Why do the quantum fluctuations of the scalar field decrease towards the origin of the time-axis? Of interest is also the inclusion of a non-zero cosmological constant. Finally, and this is most intriguing, we would like to address the question of how the quantization procedure presented here can be combined with that of the geometric sector discussed in the LQC works \cite{ashtekar:06b,ashtekar:06a,ashtekar:06c,ashtekar:07,ashtekar:11}.

\begin{acknowledgments}
AK thanks Renate Loll for discussions and Viqar Husain for reading an early version of the manuscript.
AK acknowledges support through a Projectruimte grant by the Dutch Foundation for Fundamental Research on Matter (FOM) and support by the Netherlands Organisation for Scientific Research (NWO) under their VICI program. 
TP acknowledges support by the Polish Ministerstwo Nauki i Szkolnictwa Wy\.zszego through their grant no.~182/N-QGG/2008/0, support by the Polish Narodowe Centrum Nauki through their grant no.~2011/02/A/ST2/00300 and support by the Spanish MINECO grant no.~FIS2011-30145-C03-02.
\end{acknowledgments}

\bibliographystyle{apsrev4-1}
\bibliography{archive}

\end{document}